\documentclass[12pt]{article}
\usepackage{graphicx,epsfig}
\usepackage{amsfonts,amssymb,amsmath}
\usepackage[utf8]{inputenc}
\def\eps{\varepsilon}
\def \RR {{\mathbb R}}

\def \ZZ {{\mathbb Z}}
\def \NN {{\mathbb N}}
\def \Sdot {\dot S^1}

\def\const{\mathrm{const.}}
\def\id{\mathrm{id}\,}
\def\Ad{\mathrm{Ad}\,}
\def \ket#1{{\vert #1\rangle}}
\def \HH {\mathcal{H}}
\def\rot{\mathrm{rot}}
 \def\LRA{\Leftrightarrow}
\def\inv{^{-1}}
 \def\inv{^{-1}} 
 \def\IM{\mathrm{Im}\,}
\def\ovl{\overline}

\def\be{\begin{equation}} \def\bea{\begin{eqnarray}} 
\def\ba{\begin{array}} \def\eea{\end{eqnarray}} 
\def\ee{\end{equation}}\def\ea{\end{array}}
\def \QED {\hspace*{\fill}$\square$\medskip}

\newcommand{\dd}[2]{\frac{d{#2}}{d{#1}}}
\newcommand{\qbox}[1]{\quad\hbox{#1}\quad}

\title{\bf Geometric modular action \\ for disjoint intervals \\ and boundary conformal field theory\footnote{Supported in part by ERC Advanced Grant 227458
  OACFT ``Operator Algebras and Conformal Field Theory'', and by the
  EU network ``Noncommutative Geometry'' MRTN-CT-2006-0031962. RL is
  partially supported by PRIN-MIUR and GNAMPA-INDAM. PM and KHR
  are supported in part by the German Research Foundation (Deutsche
  Forschungsgemeinschaft (DFG)) through the Institutional Strategy of
  the University of G\"ottingen.}}

\author{Roberto Longo$^{1}$, 
Pierre Martinetti$^{2,3}$, 
Karl-Henning Rehren$^{2,3}$}

\begin{document}

\maketitle

\begin{center}
\footnotesize
$^{1}$ Dipartimento di Matematica, \\
Universit\`a di Roma 2 ``Tor Vergata'', 00133 Roma, Italy \\[2mm]
$^2$ Institut f\"ur Theoretische Physik, Universit\"at G\"ottingen, \\
Friedrich-Hund-Platz 1, 37077 G\"ottingen, Germany \\[2mm]
$^3$ Courant Centre ``Higher Order Structures in Mathematics'',
  Universit\"at G\"ottingen, Bunsenstr.\ 3--5, 37073 G\"ottingen, Germany
\end{center}

\begin{abstract}
In suitable states, the modular group of local algebras associated
with unions of disjoint intervals in chiral conformal quantum field
theory acts geometrically. We translate this result into the setting
of boundary conformal QFT and interpret it as a relation between
temperature and acceleration. We also discuss aspects (``mixing'' and
``charge splitting'') of geometric
modular action for unions of disjoint intervals in the vacuum state.
\end{abstract}


\medskip

{\sl Dedicated to John E. Roberts on the occasion of his 70th birthday}

\section{Introduction}
Geometric modular action is a most remarkable feature of quantum field
theory \cite{Bo}, emerging from the combination of the basic
principles: unitarity, locality, covariance and positive energy
\cite{BW}. It associates thermal properties with localization
\cite{HHW,SW}, and is intimately related to the Unruh effect \cite{U}
and Hawking radiation \cite{S}.  It allows for a reconstruction of
space and time along with their symmetries \cite{BS}, and for a
construction of full-fledged quantum field theories \cite{KW,GLW}
out of purely algebraic data together with a Hilbert space vector (the
vacuum).  

The modular group \cite[Chap.\ VI, Thm.\ 1.19]{T} is an intrinsic
group of automorphisms of a von Neumann algebra $M$, associated with a
cyclic and separating vector $\Phi$, provided by the theory of Tomita
and Takesaki \cite{HHW,T}. In quantum field theory, $M$ may be the
algebra of observables localized in a wedge region
$\{x\in\RR^4:x^1>\vert x^0\vert\}$ and $\Phi=\Omega$ the 
vacuum state. In this situation it follows \cite{BW} that the
associated modular group is the one-parameter group of Lorentz boosts
in the $1$-direction, which preserves the wedge, i.e., it has a
geometric action on the subalgebras of observables localized in
subregions of the wedge. 

Geometric modular action was also established for the algebras of
observables localized in lightcones or double cones in the vacuum state in 
conformally invariant QFT \cite{Bu,HL}, and for interval algebras in
chiral conformal QFT \cite{GL}. It is known, however, that the 
modular group of the vacuum state is not geometric (``fuzzy'') for
double cone algebras in massive QFT (see, e.g., \cite{Bo,TS}), and the
same is true for the modular group of wedge algebras or conformal
double cone algebras in thermal states \cite{BY}. In this
contribution, we shall be interested in modular groups for algebras
associated with disconnected regions (such as unions of disjoint
intervals in chiral conformal QFT). 

Our starting point is the observation \cite{KL05} that in chiral conformal
QFT (the precise assumptions will be specified below), for any finite
number $n$ of disjoint intervals $I_i$ on the circle one can 
find states ({\em not} the vacuum if $n>1$) on the algebras $A(\bigcup_i
I_i)=\bigvee_i A(I_i)$ whose modular
groups act geometrically inside the intervals.

For $n=2$, let $E=I_1\cup I_2$ and $E'=S^1\setminus \overline E$ the
complement of the closure of $E$ . By locality, $A(E)\subset A(E')'$,
where the inclusion is in general proper. The larger algebra $A(E')'$
has the physical interpretation as a double cone algebra $B_+(O)$ in
boundary conformal QFT \cite{LR04} as will be explained in Sect.\
\ref{sec:bcft}. 

The above state on $A(E)$ can be extended to a state on $B_+(O)=
A(E')'$ such that the geometric modular action is preserved. We shall
compute the geometric flow in the double-done $O$ in Sect.\
\ref{sec:flow}. Adopting the interpretation of $\dd\tau s$ as inverse
temperature $\beta$ (where $\tau$ is the proper time along an orbit and
$s$ the modular group parameter) \cite{CR,MR}, we compute the relation
between temperature and acceleration. There is not a simple
proportionality as in the case of the Hawking temperature. 

In Sect.\ \ref{sec:CH}, we shall connect our results with a recent
work by Casini and Huerta \cite{CH}. In a first quantization approach
as in \cite{FG}, these authors have succeeded to compute the operator
resolvent in the formula of \cite{FG} for the modular operator. From
this, they obtained the modular flow for disjoint intervals and 
double cones in two dimensions in the theory of free Fermi
fields. Unlike \cite{KL05}, they consider the vacuum state. They find
a geometric modular action in the massless case (including the chiral
case), but this action involves a ``mixing'' (``modular teleportation'' 
\cite{CH}) between the different intervals resp.\ double cones. Upon
descent to gauge-invariant subtheories, the mixing leads to the new
phenomenon of ``charge splitting'' (Sect.\ \ref{sec:less}). 

Ignoring the mixing, the geometric part of the vacuum modular flow for
two intervals in the chiral free Fermi model is the same as the purely
geometric modular flow in the non-vacuum product state, provided a
``canonical'' choice for the latter is made, in the model-independent
approach.     

We shall make the result of Casini and Huerta (which was obtained by
formal manipulations of operator kernels) rigorous by establishing the
KMS property of the vacuum state w.r.t.\ modular action they found. We
shall also present a preliminary discussion of the question, to
what extent the result may be expected to hold in other than free
Fermi theories.

\section{Geometric modular flow for $n$-intervals}
\label{sec:flow}

An {\em $n$-interval} is the union $E:=\bigcup_{k=1}^n I_k$ of $n$
open intervals $I_k\subset S^1$ ($k=1,\dots,n$) with mutually disjoint 
closure. The complement $E' = S^1\setminus \ovl E$ is another
$n$-interval. If there is an interval $I\subset S^1$ such that
$E=\{z\in S^1: z^n\in I\}$, we write $E=\sqrt[n]I$, and call $E$ 
{\em symmetric}. In this case, $E'=\sqrt[n]{I'}$. Note that every
$2$-interval is a Möbius transform of a symmetric $2$-interval, while
the same is not true for $n>2$.

Let $I\to A(I)$ be a diffeomorphism covariant
local chiral net on the circle $S^1$. 
We are interested in the algebras 
\be 
  A(E):= \bigvee_{i=1}^n A(I_i)\qbox{and} \widehat A(E):= A(E')',
\ee
and their states with geometric modular action. By $\Omega$ we denote
the vacuum vector, and by $U$ the projective unitary representation of
the diffeomorphism group in the vacuum representation, with generators
$L_n$ ($n\in\ZZ$) and central charge $c$.

\subsection{Product states with geometric modular action} 
\label{sec:prodstate} 

For $n=1$, $E$ ist just an interval and $\widehat A(I)=A(I)$ (Haag
duality). 

\medskip

{\bf Proposition 1 (Bisognano-Wichmann property) \rm \cite{GL}: 
  \sl The modular group of unitaries for the pair $(A(I),\Omega)$ is
  given by the one-parameter group of Möbius transformations that
  fixes the interval $I$, $\Delta_{A(I),\Omega}^{it}=U(\Lambda_I(-2\pi t))$.} 

\medskip

For $I=S^1_+$ the upper half circle, the
generator of the subgroup $U(\Lambda_{S^1_+}(t))$ is the dilation
operator $D = i(L_1-L_{-1})$. It follows that $D$ as well as its
Möbius conjugates $D_I$ (the generators of the subgroups
$U(\Lambda_{I}(t))$) are ``of modular origin'':
\be
\label{modham}
   -2\pi \cdot D_I = \log \Delta_{A(I),\Omega} .
\ee
 
Let now 
\be 
   L_0^{(n)}= \frac 1n L_{0}+\frac c{24}\frac{n^2-1}{n},\qquad
   L_{\pm1}^{(n)} = \frac 1n L_{\pm n},
\ee
and $U^{(n)}$ the covering representation of the
Möbius group with generators $L^{(n)}_k$ ($k=0,\pm 1$). The unitary
one-parameter groups $V(t)=U^{(n)}(\Lambda_I(-2\pi t))$ act on the
diffeomorphism covariant net by 
\be
\label{nflowimpl}
   V(t) A(J) V(t)^* = A(f_t(J)) \qquad (J\subset \sqrt[n]I)
\ee
where the geometric flow $f_t$ is given by (cf.\ Fig.\ 1)
\be
\label{nflow}
   f_t(z)=\sqrt[n]{\Lambda_I(-2\pi t)(z^n)},
\ee
with the branch of $\sqrt[n]{\cdot}$ chosen in the same connected
component of $E$ as $z$, i.e., $f_t$ is a diffeomorphism
of $S^1$ which preserves each component of $E$ separately. The same
formulae hold also for $J\subset \sqrt[n]{I'}$. 

\begin{figure}[htb*]

\hskip45mm \epsfig{file=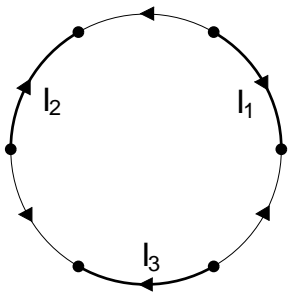,width=40mm} 

\caption{\small Flow $f_t$ in the $3$-intervals 
$E=\sqrt[3]{S^1_+}=I_1\cup I_2\cup I_3$ 
and $E'=\sqrt[3]{S^1_-}$.}

\end{figure}

The question arises whether for $n>1$ the generators $D_I^{(n)}$ of
$V(t)$ also have ``modular origin'' as in (\ref{modham}). However,
unlike with $n=1$, we have the following Lemma and Corollary: 

\medskip 

{\bf Lemma: \sl In a unitary positive-energy representation of
  $sl(2,\RR)$ of weight $h > 0$, there is no vector such that
  $D\Phi=0$, where $D=i(L_1-L_{-1})$. 

\it Proof:} An orthonormal basis of the
representation is given by the vectors $\ket n = (n!(2h)_n)^{-\frac
  12}L_{-1}^n\ket h$, where $\ket h$ is the lowest weight
vector. Solving the eigenvalue equation $L_1\Phi=L_{-1}\Phi$ by the
ansatz $\Phi=\sum_n c_n \ket n$, produces a recursion for the
coefficients $c_n$ whose solution is not square-summable.
\QED

{\bf Corollary: \sl For $n>1$, no cyclic and separating vector $\Phi$
  exists in a positive-energy representation of the net $A$ such that
  the modular Hamiltonian $\log\Delta_{A(E),\Phi}$ would equal 
  $-2\pi D_I^{(n)}$. 

\medskip 

\it Proof:} By modular theory, $\log\Delta_{A(E),\Phi}\Phi=0$. 
But because $L^{(n)}_0\geq \frac c{24}\frac{n^2-1}{n} > 0$, the Lemma
states that no vector $\Phi$ can be annihilated by $D^{(n)}_I$ which
is a Möbius conjugate of $D^{(n)}$.  
\QED

Instead, the appropriate generalization of (\ref{modham}) for the
modular origin of the generators $D_I^{(n)}$ was given in \cite{KL05},
assuming that the net $A$ is completely rational. This means that the
split property holds and the $\mu$-index $\mu_A$ is finite, and
implies the existence of a unique conditional expectation
$\eps_{E}:\widehat A(E)\to A(E)$ \cite{KLM}. In the sequel,
$\dd{\psi'}\psi$ is the Connes spatial derivative for a pair of
faithful normal states $\psi$ and $\psi'$ on a von Neumann algebra $M$
and its commutant $M'$, which is a canonical positive operator such
that $(\dd{\psi'}\psi)^{it}$ implements $\sigma^\psi_t$ on $M$ and
$(\dd{\psi'}\psi)^{-it}$ implements $\sigma^{\psi'}_t$ on $M'$
\cite{C}.  

\medskip

{\bf Proposition 2 \rm \cite{KL05}: \sl There is a faithful normal
  state $\varphi_E$ on $A(E)$ ($E=\sqrt[n]I$) and a second faithful
  normal state $\varphi_{E'}$ on $A(E')$, such that the following
  hold: The modular automorphism group $\sigma_t^{\varphi_{E}}$ is
  implemented by $V(t)$, $\sigma_t^{\varphi_{E'}}$ is implemented by
  $V(-t)$, and 
\be
\label{modori} 
  -2\pi D_I^{(n)}
  =\log\big(\frac{d\widehat\varphi_{E}}{d\varphi_{E'}}\big) + 
  \frac{n-1}2\log\mu_A.
\ee
Here, $\widehat\varphi_{E}=\varphi_{E}\circ\eps_{E}$ extends the
state on $A(E)$ to a state on $\widehat A(E)$. Moreover, 
$\frac{d\widehat\varphi_{E}}{d\varphi_{E'}} = 
\frac{d\varphi_{E}}{d\widehat\varphi_{E'}}$. }

The state $\varphi_E$ on $A(E)$ is given by 
$\varphi_E := \Big(\bigotimes_{k=1}^n \varphi_k\Big)\circ\chi_E$
where $\chi_E:A(E)\equiv \bigvee_{k=1}^nA(I_k) \to \bigotimes_{k=1}^n
A(I_k)$ is the natural isomorphism given by the split property ($I_k$
are the components of $E$), and the states $\varphi_k$ on $A(I_k)$ are
given by $\varphi_k=\omega\circ\Ad U(\gamma_k)$, where $\omega$ is the
vacuum state, and $U(\gamma_k)$ implement diffeomorphisms  
$\gamma_k$ that equal $z\mapsto z^n$ on $I_k$. (By locality,
$\varphi_k$ do not depend on the behaviour of $\gamma_k$ outside $I_k$.)

\medskip

{\bf Corollary: \sl Let $\varphi_E$ and $\widehat\varphi_E$ be the states
  on $A(E)$ and on $\widehat A(E)$, resp., as in the Proposition 2. 
For intervals $J_k\subset I_k$ (= the components of $E$) and
$F=\bigcup_{k=1}^n J_k$, we have the geometric modular actions  
\bea
\label{modflow}
  \sigma^{\varphi_E}_t(A(J_k)) = A(f_t(J_k)), \quad\hbox{hence}\quad
  \sigma^{\varphi_E}_t(A(F)) = A(f_t(F)), 
\\[2mm]
\label{dualmodflow}
  \sigma^{\widehat\varphi_E}_t(A(J_k)) = A(f_t(J_k)), \quad\hbox{and}\quad 
  \sigma^{\widehat\varphi_E}_t(\widehat A(F)) =
  \widehat A(f_t(F)). 
\eea

\it Proof:} (\ref{modflow}) is obvious from (\ref{nflowimpl}). 
By the defining implementation properties of the Connes spatial
derivative, we conclude from (\ref{modori}), that $\sigma^{\widehat\varphi_E}$ 
is implemented by $V(t)$. This implies (\ref{dualmodflow}), by the
$U^{(n)}$-covariance of the algebras under consideration. 

(We include the obvious statement (\ref{modflow}) for later comparison
with the geometric modular flow in 
\cite{CH}, for which only the second equality in (\ref{modflow}) holds
while the first is violated.)

\medskip

For $n=1$, one may just choose $\gamma=\id$, so that both $\varphi_I$ and
$\varphi_{I'}$ are given by the restrictions of the vacuum state, and
(\ref{modori}) reduces to (\ref{modham}). 

For $n>1$, the state $\varphi_E$ is different from the vacuum state,
but it is {\em rotation invariant on $A(E)$} in the sense, that 
$\varphi_E\circ \Ad{U(\rot_t)}=\varphi_E$ on $A(J_k)$ for 
$\overline J_k\subset I_k$ and $t$ small enough that
$\rot_t(J_k)\subset I_k$. ($\rot_t$ stands for the rotations $z\mapsto
e^{it}z$.) Namely, if $\ovl J\subset I$ such that $gJ\subset I$ for
$g$ in a neighborhood $N$ of the identity of the Möbius group, then by
construction, $\varphi_E\circ\Ad{U^{(n)}(g)}=\varphi_E$ on
$A(\sqrt[n]J)$ for $g\in N$. In particular, the same is true for the
rotations $\rot_t$ with $t$ in a neighborhood of $0$. Since
$U^{(n)}(\rot_t) = U(\rot_{t/n})\cdot$(complex phase), the rotation
invariance on $A(E)$ follows.

One could actually have chosen any other family of diffeomorphisms
$\gamma_k$ that map $I_k$ onto $I$, resulting in product states
$\varphi_E^{(\gamma_k)}$ with a different geometric flow on $E$. In that
case, the unitary one-parameter group $V(t)$ satisfying the properties
of the Proposition 2 is a diffeomorphism conjugate of
$U^{(n)}_I(\Lambda_I(-2\pi t))$. One might expect that our choice of
$\varphi_E$ is the only one in this class which enjoys the rotation
invariance on $A(E)$. Surprisingly, this is not the case: 

Let $\varphi_E^{(\gamma_k)}$ be a product state on $A(E)$ that is
given on $A(I_k)$ by $\omega\circ\Ad{U(\gamma_k)}$, where $\gamma_k$
are diffeomorphisms of $S^1$ that map $I_k$ onto $I$. Then this state 
is rotation invariant on $A(E)$, by construction, if and only if
$\omega\circ\Ad U(h_k)$ are rotation invariant on $A(I)$, where $h_k$
are diffeomorphisms of $S^1$, defined on $I$ by $h_k(z^n)=\gamma_k(z)$
for $z\in I_k$. In particular, $h_k$ map $I$ onto $I$. The condition
that $\omega\circ\Ad U(h)$ is rotation invariant on $A(I)$, can be 
evaluated for the two-point function of the stress-energy tensor in
that state. Using the inhomogeneous transformation law under
diffeomorphisms $h$, involving the Schwartz derivative $D_zh =
\frac{h'''}{h'} - \frac32(\frac{h''}{h'})^2$, the quantity 
\be
\label{rotinv}
  2c\cdot \left(\frac{\dd z{h_t(z)}\dd
      w{h_t(w)}}{\big(h_t(z)-h_t(w)\big)^2}\right)^2 +
  \frac{c^2}{36} \cdot D_zh_t(z) \cdot D_wh_t(w), 
\ee
where $h_t=h\circ\rot_t$, 
must be independent of $t$ for $z,w\in I$ and $t$ in a neighbourhood
of zero. Working out the singular parts of the expansion in
$w$ around $z$, one finds that $D_zh_t(z)$ must be independent of
$t$ for $z\in I$. This already implies that the second (regular) term
is separately invariant, so that, in particular, the invariance
condition does not depend on the central charge $c$. Solving 
\be
\label{schwartz}
\partial_t\big(D_zh_t(z)\big)=0 \qquad\LRA\qquad  z^2\cdot D_zh(z) =
\const, \ee
when the constant is parametrized as $\frac 12(1-\nu^2)$, 
yields
\be
\label{nu-sol}
   h(z) =\mu(z^\nu) = \frac{Az^\nu+B}{Cz^\nu+D} \quad\qbox{for} z\in I,
\ee
where $\mu$ is a Möbius transformation%
\footnote{The sign of the exponent $\nu$ can be reversed by exchanging
  $A\leftrightarrow B$ and $C\leftrightarrow D$. In order that $h$
  takes values in $S^1$, $\nu$ must be either real or imaginary, with 
  corresponding reality conditions on the matrix
  $\footnotesize{\Big(\!\ba{cc}A&B\\C&D\ea\!\Big)}$. Requiring $h$ also
  to preserve the orientation, we find: If $\nu>0$, then 
  $\footnotesize{\Big(\!\ba{cc}A&B\\C&D\ea\!\Big)}\in SU(1,1)$. If  
  $i\nu>0$, then $\scriptsize{\Big(\!\ba{cc}A&B\\C&D\ea\!\Big)}\in
  \scriptsize{\Big(\!\ba{cc}i&1\\-i&1\ea\!\Big)}\cdot SL(2,\RR)$,
  where $\scriptsize{\Big(\!\ba{cc}i&1\\-i&1\ea\!\Big)}$ 
  is the Cayley transformation $x\mapsto \frac{1+ix}{1-ix}$.}. The
state $\omega\circ \Ad U(h)$ is indeed rotation invariant on $A(I)$ by
$h\circ\rot_t(z) = \mu\circ \rot_{\nu t} (z^\nu)$ and Möbius
invariance of $\omega$. 
 
For each value of $\nu$, requiring $h$ to preserve the endpoints of the
interval $I$ fixes the Möbius transformation up to left composition
with the one-parameter subgroup $\Lambda_I(t)$. Because $\omega$ is
invariant under $\Lambda_I(t)$, the state $\omega\circ\Ad U(h)$ is
uniquely determined by the exponent $\nu$ in (\ref{nu-sol}). 

One has therefore a one-parameter family of product states, all
rotation-invariant on $A(I)$, but with different modular flows on
$I$. Going back to the product states on $A(E)$ by composition with
$z\mapsto z^n$, there is one parameter $\nu_k$ for each interval, 
i.e., for the choice of the states $\omega\circ\Ad U(\gamma_k)$ on
$A(I_k)$. The state is invariant also under ``large'' rotations by
$2\pi/n$, if and only if these parameters are the same for all $k$.

\subsection{Geometric modular action in boundary CFT}
\label{sec:bcft}

The case $n=2$ is of particular interest in boundary conformal quantum
field theory (BCFT) \cite{LR04}. With every $2$-interval $E$ such that
$-1\not\in\ovl E$, one associates a double cone $O_E$ in the halfspace 
$M_+=\{(t,x)\in\RR^2: x>0\}$ as follows. The boundary $x=0$, $t\in\RR$
is the pre-image of $\Sdot := S^1\setminus\{-1\}$ under the Cayley
transform $C:\RR\ni t\mapsto z=(1+it)/(1-it)\in S^1$. 
Let $E=I_-\cup I_+\subset \Sdot$ with $I_-<I_+$ in the counter-clockwise
order, and $I_\pm^\RR= C\inv(I_\pm)\subset\RR$. Then  
\be 
   O_E := I^\RR_+\times
I^\RR_-\equiv\{(t,x):t\pm x\in I_\pm^\RR\}.
\ee
(When there can be no confusion, we shall drop the subscript $E$.)

Now, the algebras 
\be
   B_+(O):= \widehat A(E)
\ee
have the re-interpretation as local algebras of BCFT, which extend
the subalgebras of chiral observables
\be
   A_+(O):= A(E) \equiv A(I_-)\vee A(I_+).
\ee
Under this re-interpretation, the second statement in (\ref{dualmodflow}) 
asserts, that the modular group $\sigma^{\widehat\varphi_E}_t$ acts
geometrically inside the associated diamond $O$:  
\be
  \sigma^{\widehat\varphi_E}_s(B_+(Q)) = B_+(f^O_s(Q)),
\ee
where the double cone $Q=O_F\subset O$ corresponds to a sub-2-interval
$F\subset E$, and the flow $f^O_s$ on $O$ arises from the pair of
flows $f_s$ (\ref{nflow}) on $I_+$ and $I_-$, by the said
transformations, i.e.,   
\be \label{dcflow}
f^O_s(t+x,t-x)\equiv (u_s,v_s) = (C\inv\circ f_s\circ 
C(t+x), C\inv\circ f_s\circ C(t-x)).\quad
\ee
For $I^\RR_+=(a,b)\subset \RR_+$ and $I^\RR_-=(-1/a,-1/b)$ (corresponding 
to a symmetric $2$-interval $E$), we have computed the velocity field 
\be 
\label{symdiff}
  \partial_s u_s = 2\pi \frac{(u_s-a)(au_s+1)(u_s-b)(bu_s+1)}
  {(b-a)(1+ab)\cdot(1+u_s^2)} =: -2\pi V^O(u_s)
\ee
for $u_s\in I^\RR_+$, and the same equation for $v_s\in I^\RR_-$.

For $I^\RR_+=(a_1,b_1)$ and $I^\RR_-=(a_2,b_2)$ corresponding to a
non-symmetric $2$-interval $\tilde E$, there is a Möbius
transformation $m$ that maps $\tilde E$ onto a symmetric interval
$E$. Choosing the state $\varphi_{\tilde E}:=\varphi_{E}\circ\Ad U(m)$
on $A(\tilde E)$, the resulting geometric modular flow is given by
$\tilde f_s = m\inv \circ f_s\circ m$. Going through the same steps,
we find  
\be 
\label{diff}
\partial_s u_s = -2\pi V^O(u_s) = 
2\pi \frac{(u-a_1)(u-b_1)(u-a_2)(u-b_2)}{Lu^2-2Mu+N}
\ee
with
\be
L=b_1-a_1+b_2-a_2,\quad M= b_1b_2-a_1a_2,\quad
N=b_2a_2(b_1-a_1)+b_1a_1(b_2-a_2). \quad
\ee
This differential equation is solved
by 
\be
\label{sol}
  \log-\frac{(u_s-a_1)(u_s-a_2)}{(u_s-b_1)(u_s-b_2)}= -2\pi s + \const
\ee
The modular orbits for $u=t+x,v=t-x$ are obtained by eliminating $s$:
\be
\label{orb}
  \frac{(u-a_1)(u-a_2)}{(u-b_1)(u-b_2)}\cdot
  \frac{(v-b_1)(v-b_2)}{(v-a_1)(v-a_1)}=\const
\ee

\subsection{General boundary CFT}

Up to this point, we have taken the boundary CFT to be given by 
$B_+(O):=\widehat A(E)$, which equals the relative commutant
$B_+(O)= A(K)'\cap A(L)$ by virtue of Haag duality of the local chiral
net $A$. Here, $K$ and $L\subset\Sdot$ are the open
intervals between $I_+$ and $I_-$, and spanned by $I_+$
and $I_-$, respectively, i.e., $L=I_+\cup \ovl K\cup I_-$. 

The general case of a boundary CFT was studied in \cite{LR04}. If $A$
is completely rational, every irreducible local boundary CFT net
containing $A(E)$ is intermediate between $A(E)$ and a maximal (Haag
dual) BCFT net: 
\be
  A(I_+)\vee A(I_-) \equiv A_+(O) \subset B_+(O)\subset B_+^{\rm dual}
  (O)\equiv B(K)'\cap B(L),
\ee
where $I\mapsto B(I)$ is a conformally covariant, possibly nonlocal
net on $\Sdot$ (its extension to the circle in general requires a
covering), which extends $A$ and is relatively local w.r.t.\ $A$. If $A$
is completely rational, the local subfactors $A(I)\subset B(I)$
automatically have finite index (not depending on $I\subset\Sdot$), and
there are only finitely many such extensions.  

There is then a unique global conditional expectation $\eps$, that maps
each $B(I)$ onto $A(I)$. $\eps$ commutes with Möbius transformations
and preserves the vacuum state. By relative locality, $\eps$ maps  
$B(K)'\cap B(L)$ into (in general not onto) $A(K)'\cap A(L)$, hence  
\be
  A(E)\equiv  A_+(O)\subset \eps(B_+(O)) \subset \widehat A(E).
\ee
The product state $\widehat\varphi_E$ on $\widehat A(E)$ 
induces a faithful normal state $\widehat\varphi_E\circ\eps$ on 
$B_+(O)$. 

\medskip

{\bf Proposition 3: \sl In a completely rational, diffeomorphism
  invariant BCFT, the modular group of the state
  $\widehat\varphi_E\circ\eps$ acts geometrically on $B_+(Q)$,
  $Q\subset O$, i.e., $\sigma^{\widehat\varphi_E\circ\eps}_s(B_+(Q)) =
  B_+(f^O_s(Q))$, where $f^O_s$ is the flow (\ref{dcflow}). 

\medskip 

\it Proof:} $B_+(O)$ is generated by $A_+(O)$ and an isometry $v$
\cite{LR04} such that every element $b\in B_+(O)$ has a unique
representation as $b=av$ with $a\in A_+(O)$, and $va=\theta(a)v$ where
$\theta$ is a dual canonical endomorphism of $B_+(O)$ into $A_+(O)$. 
For a double cone $Q\subset O$, the isometry $v$ may be choosen to
belong to $B_+(Q)$, in which case $\theta$ is localized in $Q$. We
know that the modular group restricts to the modular group of
$A_+(O)$, which acts geometrically, in particular, it takes $A_+(Q)$
to $A_+(f^O_s(Q))$. It then follows by the properties of the
conditional expectation that $\sigma^{\widehat\varphi_E\circ\eps}_s(v)
\equiv v_s = u_sv$ where $u_s\in A(E)$ is a unitary cocycle of
intertwiners $u_s:\theta\to\theta_s\equiv \sigma^{\widehat\varphi_E}_s
\circ\theta\circ\sigma^{\widehat\varphi_E}_s{}\inv$. Since  
$\sigma^{\widehat\varphi_E}_s$ acts geometrically in $A_+(O)$,
$\theta_s$ is localized in $f^O_s(Q)$, and $A_+(f^O_s(Q))\cdot v_s =
B_+(f^O_s(Q))$. This proves the claim. 
\QED

Thus, in every BCFT, the modular group of the state
$\widehat\varphi_E\circ\eps$ on $B_+(O_E)$ acts geometrically inside
the double cone $O_E$ by the same flow (\ref{sol}), (\ref{orb}).

\subsection{Local temperature in boundary conformal QFT} 
\label{sec:localtemp}

We shall show that the states $\widehat\varphi_E\circ\eps$, whose
geometric modular action we have just discussed, are manufactured
far from thermal equilibrium. We adopt the notion of ``local
temperature'' introduced in \cite{BOR}, where one compares the
expectation values of suitable ``thermometer observables'' $\Phi(x)$ 
in a given state $\varphi$ with their expectation values in global KMS
reference states $\omega_\beta$ of inverse temperature $\beta$. If one
can represent the expectation values as weighted averages   
\be
\varphi(\Phi(x)) = \int d\rho_x(\beta)\, \omega_\beta(\Phi(x))
\ee
(where the thermal functions $\beta\mapsto \omega_\beta(\Phi(x))$ do
not depend on $x$ because KMS states are translation invariant), then
one may regard the state $\varphi$ at each point $x$ as a statistical
average of thermal equilibrium states. In BCFT, this analysis can
be carried out very easily for the product states $\varphi_E$ with the
energy density $2T_{00}(t,x)=T(t+x)+T(t-x)$ as thermometer
observable. One has $\omega_\beta(T(\,\cdot\,)) = 
\frac{\pi^2}{24}c\,\beta^{-2}$ in the KMS states, while the
inhomogeneous transformation law of $T$ under diffeomorphisms gives
$\varphi_E(T(y)) = -\frac c{24\pi}\, D_y\gamma_\pm(y) = 
-\frac c{4\pi} \, (1+y^2)^{-2}$ if $y\in I^\RR_\pm$ where
$\gamma_\pm(y) = C\inv\circ (z\mapsto z^2)\circ C(y)=\frac{2y}{1-y^2}$, 
i.e., negative energy density inside the double cone $O=I^\RR_+\times
I^\RR_-$. The product states $\varphi_E$ can therefore not be interpreted 
as local thermal equilibrium states in the sense of \cite{BOR}. The
possibility of locally negative energy density in quantum field theory
is well-known, and its relation to the Schwartz derivative in
two-dimensional conformal QFT was first discussed in \cite{F}.

\subsection{Modular temperature in boundary conformal QFT} 
\label{sec:modtemp}

The ``thermal time hypothesis'' \cite{CR} provides a very different
thermal interpretation of states with geometric modular action. 
According to this hypothesis, one interprets the norm of the vector
$\partial_s$ tangent to the modular orbit $x^\mu(s)$ as the inverse
temperature $\beta_s$ of the state as seen by a physical observer with
accelerated trajectory $x^\mu(s)$. In the vacuum state on the Rindler
wedge algebra, this gives precisely the Unruh temperature $\beta_s=\dd
s\tau =\frac{2\pi}{\kappa}$ ($\tau$ is the proper time, and $\kappa$
the acceleration). One may also give a local interpretation, by viewing
$\beta_s$  as the inverse temperature of the state for an observer at
each point whose trajectory is tangent to the unique modular orbit
through that point.

For these interpretations to make sense it is important that
$\partial_s$ is a timelike vector. Indeed, it is easily seen that the
flow (\ref{symdiff}), (\ref{diff}) gives negative sign for both
$\partial_s u_s$ and $\partial_s v_s$, because the velocity field
$V^O$ is positive inside the interval. Hence the tangent vector is
past-directed timelike. This conforms with a general result, proven in 
{\em more than 2} spacetime dimensions:  

\medskip

{\bf Proposition 4 \rm \cite{ST}: \sl Let $A(O)$ be a local algebra
  and $U_t$ a unitary one-parameter group such that 
  $U_tA(Q)U_t^*=A(f_tQ)$ where $f_t$ is an automorphism of $O$ taking
  double cones in $O$ to double cones. If there is a vector $\Phi$,
  cyclic and separating for $A(O)$, such that $U_tA\Phi$ has an
  analytic continuation into a strip $-\beta < \IM t < 0$, then
  $-\partial_t(f_tx)\vert_{t=0}\in \overline{V_+}$. In particular, the
  flow of a geometric modular action is always past-directed null or
  timelike.}  

\medskip
 
From (\ref{diff}), we get the proper time $(d\tau)^2 = du\,dv$ and
hence the inverse temperature $\beta=\dd s\tau$ as a function of the
position $x^\mu=(t,x)$  
\be 
\label{beta}
  \beta(t,x)^2=\frac {du}{ds}\frac {dv}{ds} = 4\pi^2\cdot
  V^O(t+x)V^O(t-x).
\ee
The temperature diverges on the boundaries of the double cone
($V^O(a_i)=V^O(b_i)=0$), and is positive everywhere
in its interior.

For comparison with the ordinary Unruh effect, we also compute the
acceleration in the momentarily comoving frame
\be
   \kappa = \Big(-\frac{\partial^2x_\mu}{\partial\tau^2}
   \frac{\partial^2x^\mu}{\partial\tau^2}\Big)^{\frac12}=
   \frac {(d^2x/dt^2)}{(1-(dx/dt)^2)^{3/2}} = 
   \frac{u''v'-u'v''}{2(u'v')^{3/2}},
\ee 
where the prime stands for $\partial_s$, and we have used $\dd tx =
\frac{x'}{t'} = \frac {u'-v'}{u'+v'}$ and $\frac {d^2x}{dt^2} = 
\frac {(dx/dt)'}{t'} = 4\frac{u''v'-u'v''}{(u'+v')^3}$. Thus    
\be
   \kappa(t,x) =
   \frac{V^O{}'(u)-V^O{}'(v)}{2\sqrt{V^O(u)V^O(v)}}\Big\vert_{u=t+x,\,v=t-x} 
   = \frac {V^O{}'(t+x)-V^O{}'(t-x)}{\pi\inv\beta(t,x)} 
\ee
as a function of the position $(t,x)$. The product  
\bea
  \beta(t,x)\cdot \kappa(t,x) &=& \pi \big\vert\partial_x 
  \big(V^O(t+x)+V^O(t-x)\big)\big\vert \nonumber \\ 
  &=& \pi \big\vert\partial_t \big(V^O(t+x)-V^O(t-x)\big)\big\vert\quad
\eea
has the maximal value $2\pi$ (Unruh temperature) near the left and
right edges of the double cone, and equals $0$ along a timelike curve
connecting the past and future tips. This curve is in general not
itself a modular orbit.

In general, the modular orbits are not boost trajectories. However,
the quantitative departure is very small. As an illustration, we
display a true modular orbit, as well as a plot with one coordinate
exaggerated by a zoom factor of 100 (Fig.\ 2).

\begin{figure}[htb*]
\hskip18mm \epsfig{file=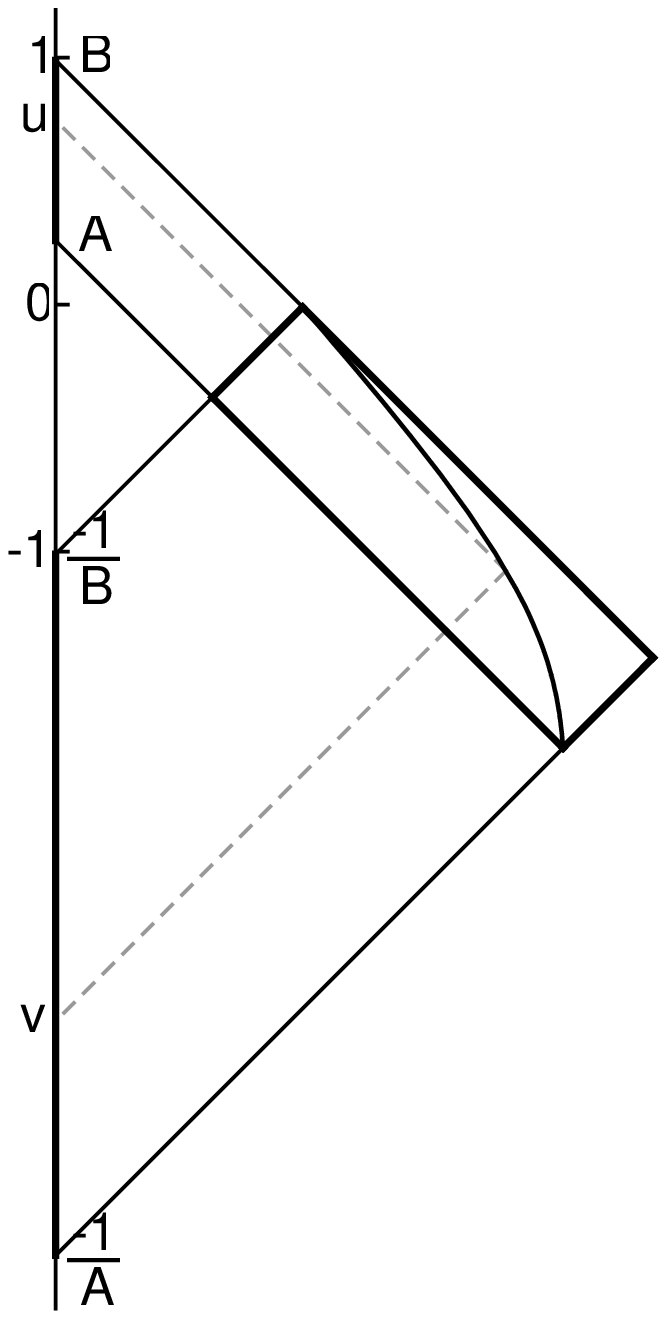,width=52mm} 
\epsfig{file=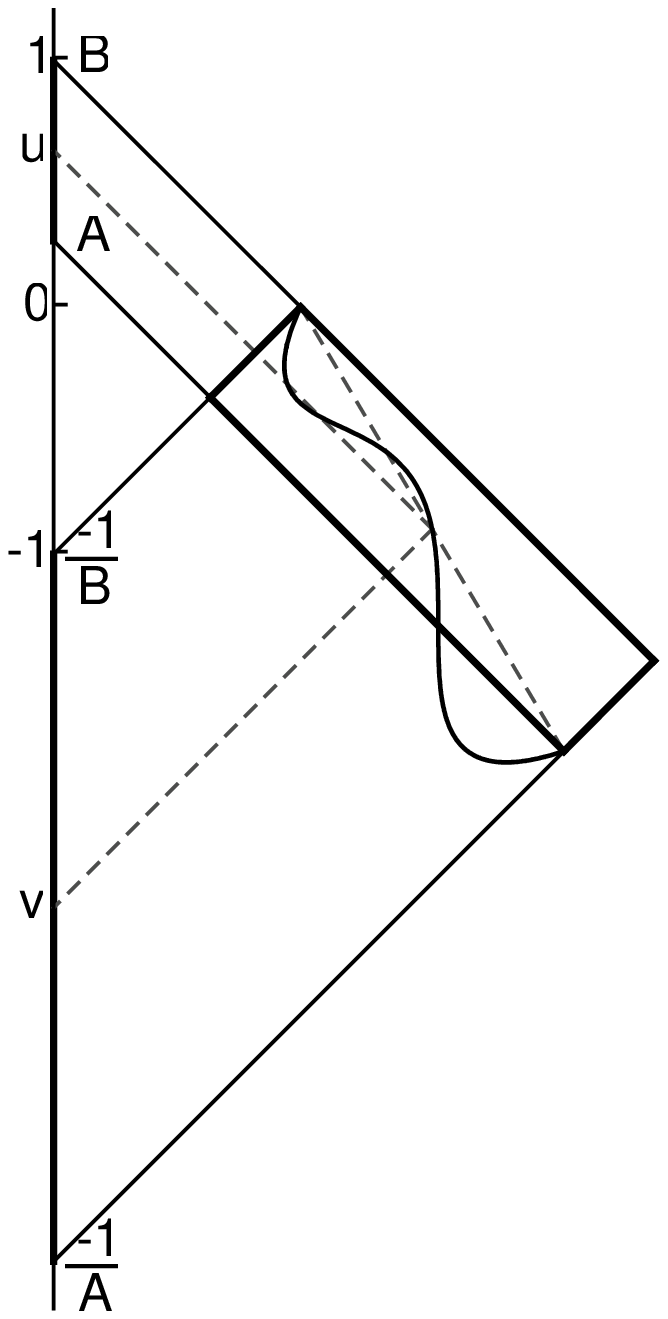,width=52mm} 
\caption{\small Influence of the boundary. Left: modular orbit of an
  arbitrary point in the symmetric double cone $O=\{(t,x):A\leq t+x\leq
  B, -\frac1A\leq t-x\leq -\frac1B\}$. Right: a zoom on the modular
  orbit $(u_s, v_s)$ going through the center of the double cone. The
  plot represents the curve $(\tilde{u}_s,v_s)$ where $\tilde{u}_s =
  f*(u_s - u^{\text{diag}}_s) +u^{\text{diag}}_s$, with
  $(u^{\text{diag}}_s, v_s)$ the straight line joining the two tips of
  the double cone (a special vacuum modular orbit in the absence of
  the boundary), and $f=100$ a zoom factor.} 
\end{figure}

There exists however one distinguished modular orbit with a simple
dynamics, namely the boost  
\be
\label{boost}
  u_s v_s= -1  \qquad \forall s\in\RR 
\ee
(in the symmetric case, for simplicity) which is solution of
(\ref{orb}) for $\const =1$. It is the Lorentz boost of a wedge in
$M_+$, whose edge lies on the boundary $x=0$. The same is true also
for non-symmetric intervals, although the formula (\ref{boost}) is
more involved.

Along this distinguished orbit the inverse temperature (\ref{beta})
simply writes  
\be
\label{eq:15}
  \beta = 2\pi \frac{\partial_s u_s}{u_s} = 2\pi\frac{d}{ds}\ln u_s. 
\ee
One can express the proper time $\tau$ of the observer following the
boost as a function of the modular parameter 
\be
   \tau(s) = \ln u_s - \ln u_0,
\ee
hence $\beta(\tau) = 2\pi\frac{V^O(u_0 e^\tau)}{u_0 e^\tau}$. Choosing
$u_0=1$, one can write the inverse temperature as a function of the
proper time in the form
\be
   \beta(\tau)  =  2\pi\frac{(\sinh(\tau_{\rm  max})-\sinh(\tau))\cdot 
  (\sinh(\tau)-\sinh(\tau_{\rm min}))}
  {(\sinh(\tau_{\rm max})-\sinh(\tau_{\rm min}))\cdot \cosh(\tau)},
\ee
where $\tau_{\rm min}$ and $\tau_{\rm max}$ are functions of the
coordinates of the double cone. As for double cones in Minkowski
space, the temperature is infinite at the tips of the double cone
($\tau=\tau_{\rm min}$ or $=\tau_{\rm  max}$) and reaches its minimum
in the middle of the observer's ``lifetime''. Unfortunately, for
generic orbits we have no closed formula for the temperature as a
function of the proper time, so as to compare with the ``plateau
behaviour'' (constant temperature for most of the ``lifetime'') as in
\cite{MR}, that occurs in CFT without boundary for vacuum modular
orbits close to the edges of the double cone.  

\section{The Casini-Huerta modular flow}
\label{sec:CH}
\setcounter{equation}0

Casini and Huerta recently found \cite{CH} that the vacuum modular
group for the algebra of a free Fermi field in the union of $n$
disjoint intervals $(a_k,b_k)\subset\RR$ is given by the formula
\be
\label{CHflow}
  \sqrt{\textstyle\frac{dx_j}{d\zeta}}\cdot \sigma_t(\psi(x_j)) =
  \sum_{k} O_{jk}(t) \sqrt{\textstyle\frac{dx_k(t)}{d\zeta}} \cdot
  \psi(x_k(t)).
\ee 
Here, 
\be
\label{zeta} 
   e^{\zeta(x)}=-\prod_k\frac{x-a_k}{x-b_k}
\ee
defines a uniformization function $\zeta$ that maps each interval
$(a_k,b_k)$ onto $\RR$, and $e^\zeta\in\RR_+$ has $n$ pre-images
$x_k=x_k(\zeta)$, one in each interval, i.e.,
$-\prod_l\frac{x_k(\zeta)-a_l}{x_k(\zeta)-b_l}=e^\zeta$. The geometric
modular flow is given by%
\footnote{In \cite{CH}, the notation is different: the authors
  ``counter'' the flow so that the position of
  $\sigma_t(\psi(x_j(\zeta+2\pi t)))$ remains constant, except for the
  mixing.} 
\be
  \label{zeta(t)}
\zeta(t)=\zeta_0-2\pi t,
\ee
i.e., a separate flow $x_k(t)=x_k(\zeta-2\pi t)$ in each interval. The
orthogonal matrix $O$ yields a ``mixing'' of the fields on the different 
trajectories $x_i(t)$, and is determined by the differential equation 
\be
\label{mixdiff} 
   \dot O(t) = K(t)O(t) 
\ee
where $K_{jj}(t)=0$ and 
\be
   K_{jk}(t) = 2\pi \frac{\sqrt{\frac{dx_j(t)}{d\zeta}}
   \sqrt{\frac{dx_k(t)}{d\zeta}}}{x_j(t)-x_k(t)} \qquad
   (j\neq k).
\ee

{\bf Remark: \sl The mixing is a ``minimal'' way to evade an
  absurd conclusion from Takesaki's Theorem \cite[Chap.\ IX, Thm.\
  4.2]{T}: Without mixing the modular group would globally preserve
  the component interval subalgebras. Then the Reeh-Schlieder
  property of the vacuum vector would imply that the $n$-interval
  algebra coincides with each of its component interval subalgebras.} 

\medskip

{\bf Proposition 5: \sl For $\bigcup_{k}(a_k,b_k)\subset \RR$ the
  Cayley transform of a symmetric $n$-interval $E=\sqrt[n] I\subset
  S^1\setminus\{-1\}$, the geometric part (\ref{zeta(t)}) of the flow
  (without mixing) is the same as (\ref{nflow}).  

\medskip

\it Proof:} We use variables $u_k=\frac{1+ia_k}{1-ia_k}$,
$v_k=\frac{1+ib_k}{1-ib_k}$, $z=\frac{1+ix}{1-ix}$, and the identity
$2i(x-a) = (1-ix)(1-ia)(z-u)$. Then
\be 
  e^\zeta = -\prod_k\frac{x-a_k}{x-b_k} =
  \const\cdot\prod_k\frac{z-u_k}{z-v_k} = \const\cdot
  \frac{z^n-U}{z^n-V}
\ee
where $U=u_k^n$, $V=v_k^n$ such that $I=(U,V)\subset S^1$. 
Therefore, the flow (\ref{zeta(t)}) is equivalent to 
\be
\label{zt} 
  \frac{z(t)^n-U}{z(t)^n-V}= e^{-2\pi t}\cdot\frac{z^n-U}{z^n-V},
\ee
which in turn is easily seen to be equivalent to (\ref{nflow}). 
\QED

Keep in mind, however, that the modular group of the product state in
Sect.\ \ref{sec:prodstate} does not ``mix'' the intervals $(a_k,b_k)$. 

Since every 2-interval is a Möbius transform of a symmetric
2-interval, the statement of Prop.\ 5 is also true for general
2-intervals, with the flow (\ref{sol}).

\subsection{Verification of the KMS condition}

The authors of \cite{CH} have obtained the flow (\ref{CHflow}) using
formal manipulations. We shall establish the KMS property of the
vacuum state for this flow. Because this property distinguishes the
modular group \cite{T}, we obtain an independent proof of the claim. 

We take $\bigcup_{k}(a_k,b_k)\subset \RR$ the Cayley transform
of a symmetric $n$-interval $E=\sqrt[n] I\subset \Sdot$. 
We first solve the differential equation (\ref{mixdiff}) for the
mixing. 

With angular variables $x=\tan\frac\xi2$, and $\pi > \xi_0 > \xi_1 >
\cdots > \xi_{n-1} > -\pi$, the non-diagonal elements of the matrix
$K$ can be written as 
\be
  K_{kl}(t) = 2\pi\cdot \frac{\sqrt{\dd {\xi_k(t)}{x_k(t)}} 
  \sqrt{\dd {\xi_l(t)}{x_l(t)}}}{x_k(t)-x_l(t)} 
  \sqrt{\textstyle\dd  z{\xi_k(t)}}  
  \sqrt{\textstyle\dd z {\xi_l(t)}} = 2\pi\cdot \frac{\sqrt{\dd z{\xi_k(t)}} 
  \sqrt{\dd z {\xi_l(t)}}}{2\sin\frac{\xi_k(t)-\xi_l(t)}2} \quad
\ee
for $k\neq l$. For symmetric intervals, 
$\xi_k= \xi_0-k\cdot\frac{2\pi}n$ and $\dd z{\xi_k} = \dd
z{\xi_0} >0$, hence 
\be
\label{omega}
  K_{kl}(t)=- 2\pi\cdot \frac{\dd z {\xi_0(t)}}{2\sin\frac{(k-l)\pi}n}
  = \Omega_{kl}\cdot \dot \xi_0(t), \qquad \Omega_{kl}=
  \frac{1}{2\sin\frac{(k-l)\pi}n}. 
\ee
With $\Omega=(\Omega_{kl})_{k,l=0}^{n-1}$ the constant matrix with
vanishing diagonal elements, we obtain the orthogonal mixing matrix 

\medskip 

{\bf Corollary: \sl The mixing matrix is given by} 
\be 
   O(t)=e^{(\xi_0(t)-\xi_0(0))\cdot \Omega}.
\ee

{\bf Remark: \sl The mixing matrix $O(t)$ always belongs to the same
  one-parameter subgroup of $SO(n)$, with generator $\Omega$. For
  $n=2$, this is just 
\be 
  O(t)=\Big(\ba{cc}\cos\theta & -\sin\theta \\
  \sin\theta & \cos\theta 
\ea\Big) \qquad\hbox{with}\quad \theta(t)=\frac 12(\xi_0(t)-\xi_0).
\ee
If $E$ is not symmetric, the general formula is 
\be
\label{angle}
  \theta(t) = \arctan\frac{Lx_0(t)-M}{\sqrt{LN-M^2}} -
  \arctan\frac{Lx_0(0)-M}{\sqrt{LN-M^2}}
\ee
with notations as in (\ref{diff}).}\footnote{The authors of \cite{CH}
  also compute this angle, but misrepresent it as the $\arctan$ of the
  difference, rather than the difference of the $\arctan$'s.}

\medskip

Next, we compute the vacuum expectation values
$\langle\sigma_t(\psi(x_i))\sigma_s(\psi(y_j))\rangle$, using
(\ref{CHflow}) and $\langle\psi(x)\psi(y)\rangle
=\frac{-i}{x-y-i\eps}$. Passing to angular variables $x\mapsto \xi$,
$y\mapsto\eta$ by 
\be
\label{angular}
   \frac{\sqrt {dx}\sqrt{dy}}{x-y-i\eps} = 
  \frac {\sqrt{d\xi}\sqrt{d\eta}}{2\sin\frac{\xi-\eta-i\eps}2},
\ee
this gives
\bea
\label{vev1} 
  \langle\sigma_t(\psi(x_i))\sigma_s(\psi(y_j))\rangle =
  \hspace{70mm}\nonumber \\[-4mm]
  =\sum_{kl}\Big(e^{(\xi_0(t)-\xi_0)\cdot \Omega}\Big)_{\!ik}
  \Big(e^{(\eta_0(s)-\eta_0)\cdot \Omega}\Big)_{\!jl}\cdot
  \frac{-i\sqrt{\dd{x_i}{\xi_k(t)}}\sqrt{\dd{y_j}{\eta_l(s)}}}
  {2\sin(\frac{\xi_k(t)-\eta_l(s)-i\eps}2)}.  
\eea
Notice that again $d\xi_k$, $d\eta_l$ in the square roots do not
depend on $k$ and $l$. To perform the sums over $k$ and $l$, we need a
couple of trigonometric identities:

\medskip

{\bf Lemma: \sl For $n\in\NN$ and $k=0,1,\dots n-1$, let
  $\sin_k(\alpha):=\sin(\alpha-k\frac \pi n)$. Then 
 (sums and products always extending from $0$ to $n-1$):

(i) $ \prod_{k} \sin_k(\alpha) = (-2)^{1-n}\sin(n\alpha)$.

(ii) For $j=0,\dots,n-1$ one has $\sum_{k:\,k\neq j}
\cot((j-k)\frac\pi n) = 0$. 

(iii) For $j=0,\dots,n-1$ one has
\be
\label{lemma} 
  \sum_{k} \Big(e^{2(\alpha-\beta)\Omega}\Big)_{jk}\cdot 
  \frac1{\sin_k(\alpha)} = \frac{\sin(n\beta)}{\sin(n\alpha)} \cdot 
  \frac 1{\sin_j(\beta)}.
\ee

\medskip

\it  Proof:} (i) is just another way of writing 
$\prod_k (z-\omega_k) = z^n-1$ where $\omega_k=e^{ik\frac {2\pi}n}$
are the $n$th roots of unity, and $z=e^{2i\alpha}$. Dividing
(i) by $\sin_j(\alpha)$, taking the logarithm, and taking
the derivative at $\alpha=0$, yields (ii). For (iii), we have to show
that the expression  
\be
  (-2)^{1-n}\sin(n\alpha)\sum_{k} \Big(e^{2\alpha\Omega}\Big)_{jk}
  \cdot \frac 1{\sin_k(\alpha)} =
  \Big(e^{2\alpha\Omega}\Big)_{jk}
  \cdot \prod_{l:\;l\neq k} \sin_l(\alpha) 
\ee
is independent of $\alpha$. Taking the derivative w.r.t.\ $\alpha$ and
inserting (\ref{omega}), we have to show that 
\be
\sum_{k} \frac 1{\sin(j-k)\frac \pi n} \cdot
\prod_{l:\;l\neq k} \sin_l(\alpha) + \sum_{k} 
\cos(\alpha - k\frac \pi n)\cdot\prod_{l:\;l\neq j,k}
\sin_l(\alpha) =0.\quad
\ee
Writing $ \cos(\alpha - k\frac \pi n)=\big(
\sin_k(\alpha)\cos ((j-k)\frac\pi n)-\sin_j(\alpha)\big)/\sin
((j-k)\frac\pi n)$, this sufficient condition reduces to the identity (ii). 
\QED

Using (\ref{lemma}) with $2\alpha=\xi_0(t)-\eta_l(s)$ and
$2\beta=\xi_0-\eta_l(s)$ in the expression (\ref{vev1}), and once again
with $2\alpha=\eta_0(s)-\xi_0$ and
$2\beta=\eta_0-\xi_0$, we get 
\be
\label{vev2} 
  \langle\sigma_t(\psi(x_i))\sigma_s(\psi(y_j))\rangle
  =\frac{\sin(n\frac {\xi_0-\eta_0-i\eps}2)}
  {\sin(n\frac{\xi_0(t)-\eta_0(s)-i\eps}2)}\frac
  {-i\sqrt{\dd{x_i}{\xi_0(t)}}\sqrt{\dd{y_j}{\eta_0(s)}}} 
  {2\sin(\frac{\xi_i-\eta_j-i\eps}2)}.
\ee
We exhibit the $t$- and $s$-dependent terms:
\be
  \frac{\sqrt{d\xi_0(t)}\sqrt{d\eta_0(s)}} 
       {2\sin(\frac {n\xi_0(t)-n\eta_0(s)-i\eps}2)} = 
  \frac{\sqrt{d\Xi_0(t)}\sqrt{d\mathrm{H}_0(s)}} 
       {2n\sin(\frac {\Xi_0(t)-\mathrm{H}_0(s)-i\eps}2)} = 
  \frac 1n \frac {\sqrt {dX(t)}\sqrt{dY(s)}} {X(t)-Y(s)-i\eps}. 
\ee
The first equality is the invariance of the 2-point function under a
Möbius transformation $\mu$ mapping $I$ to $S^1_+$, such that for
$z=e^{i\xi}\in E$ and $w=e^{i\eta}\in E$ we get $\mu(z^n) =
e^{i\Xi}=\frac{1+iX}{1-iX}\in S^1_+$ and $\mu(w^n) =e^{i\mathrm{H}} = 
\frac{1+iY}{1-iY}\in S^1_+$ with $X,Y\in\RR_+$; the second equality 
is again (\ref{angular}) for the inverse transformation
$\Xi\mapsto X$, $\mathrm{H}\mapsto Y$. By Prop.\ 5, the flow on
$\RR_+$ is just $X(t)=e^{-2\pi t}\cdot X$, giving
\be 
\label{vev3}
  \langle\sigma_t(\psi(x_i))\sigma_s(\psi(y_j))\rangle =  
  \frac {e^{-\pi (t+s)}}{e^{-2\pi t}X-e^{-2\pi s}Y-i\eps}\cdot
  f(x_i,y_j).
\ee
This expression manifestly satisfies the KMS condition in the form 
\be 
\label{kms}
   \langle\psi(x)\sigma_{-i/2}(\psi(y))\rangle = 
   \langle\psi(y)\sigma_{-i/2}(\psi(x))\rangle.
\ee
We conclude that
the KMS condition holds for the Casini-Huerta flow for symmetric
$n$-intervals: 

\medskip

{\bf Corollary: \sl For symmetric $n$-intervals $E=\sqrt[n]I$,
  (\ref{CHflow}) is the modular automorphism group of the algebra
  $A(E)$ with respect to the vacuum state. 

\medskip

\it Proof:} Smearing with test functions of appropriate support, the
KMS property holds for bounded generators of the CAR algebra $A(E)$. 
Because $\psi$ is a free field, the KMS property of the 2-point
function in the vacuum extends to the KMS property of the
corresponding quasifree (i.e., Fock) state of the CAR algebra. \QED 

\medskip

{\bf Remark: \sl It is quite remarkable that by virtue of the mixing,
  through the identity (ii) of the Lemma, the ratio of the modular
  vacuum correlation functions
\be
  \frac{\langle\sigma^{(n)}_t(\psi(x_i))\sigma^{(n)}_s(\psi(y_j))\rangle}
  {\langle\sigma^{(1)}_t(\psi(X))\sigma^{(1)}_s(\psi(Y))\rangle}
\ee
is independent of the modular parameters $t,s$. Here, in the
numerator $\sigma^{(n)}$ is the modular group for a symmetric
$n$-interval $\subset \RR$, and in the denominator $\sigma^{(1)}$ is
the modular group for the $1$-interval $\RR_+$.}

\subsection{Product states for general $n$-intervals} \label{sec:genprod}

With hindsight from \cite{CH}, we can generalize to non-symmetric
$n$-intervals the model-independent construction of a product state,
as in Sect.\ \ref{sec:prodstate}, by replacing the function 
$z\mapsto z^n$ as follows. If $C$ stands for the Cayley transformation
$x\mapsto z= 
\frac{1+ix}{1-ix}$, and $\bigcup_k(a_k,b_k)\subset\RR$ the pre-image of
a symmetric $n$-interval $E=\sqrt[n]I$, then $U=C(a_k)^n\in S^1$ and
$V=C(b_k)^n\in S^1$ do not depend on $k$. One computes the
uniformization function (\ref{zeta}) in this case to be given by 
\be 
   e^\zeta = C\inv\circ\mu\circ (z\mapsto z^n) \circ C(x)
\ee
where $\mu:S^1\to S^1$ is the Möbius transformation $Z\to
C\Big(\frac{(-1)^n-V}{(-1)^n-U}\cdot\frac{Z-U}{V-Z}\Big)$,
that takes $I$ to $S^1_+$. For a general $n$-interval 
$E=\bigcup I_k\subset \Sdot$, one may choose $\mu$ an
arbitrary Möbius transformation, and replace $z\mapsto z^n$ by the
function  
\be 
\label{g} 
  g(z):= \mu\inv\circ C\circ e^\zeta\circ C\inv,
\ee
where $\zeta$ is the uniformization function (\ref{zeta}). Thus, $g$
maps each component $I_k$ onto the same interval $I=\mu\inv(S^1_+)$,
i.e., we have $E=g\inv(I)$. Repeating the construction of Prop.\ 2
with factor states $\varphi_k=\omega\circ\Ad U(\gamma_k)$, where the
diffeomorphisms $\gamma_k$ coincide with $g$ on $I_k$, one obtains a
product state with the geometric modular flow 
\be
   f_t(z)=g\inv\big(\Lambda_I(-2\pi t)g(z)\big),
\ee
instead of (\ref{nflow}). By construction, this flow corresponds to
$\zeta(t)=\zeta(0)-2\pi t$ as before, which in turn coincides with the
geometric part of the vacuum modular flow (\ref{CHflow}). 
 
\subsection{Lessons from the free Fermi model}
\label{sec:less}

{\bf Charge splitting}

It is tempting to ask whether, and in which precise sense, the free
Fermi field result extends also to the free Bose case. (The authors of
\cite{CH} are positive about this, but did not present a proof.) In
the chiral situation, the free Bose net $A(I)$ (the current algebra
with central charge $c=1$) is given by the neutral subalgebras of the
complex free Fermi net $F(I)$. Because the vacuum state is invariant
under the charge transformation, there is a vacuum-preserving
conditional expectation $\eps:F(I)\to A(I)$, implying that the vacuum
modular group of $F(E)$ restricts to the vacuum modular group of
$C(E):=\eps(F(E))$. We have 
\be
\label{incl}
   \ba{ccccc}&& F(E)&& \\ &&\eps\downarrow\quad&& \\ A(E)&\subset
   & C(E)& \subset & \widehat A(E),\ea 
\ee
where both inclusions are strict: $C(E)$ contains neutral products of
integer charged elements of $F(I_k)$ in different component intervals,
which do not belong to $A(E)$, while $\widehat A(E)$ contains ``charge
transporters'' \cite{BMT,KLM} for the continuum of superselection
sectors of the current algebra with central charge $c=1$, which do not
belong to $C(E)$.  

Being the restriction of the vacuum modular group of $F(E)$, 
the action of the vacuum modular group of $C(E)$ can be directly read
off. It acts geometrically, i.e., takes 
$C(F)$ to $C(f_t(F))$,\footnote{Here and below, $F\subset E$ always
  stands for an $n$-interval $F=\bigcup_k J_k$ where $J_k$ are the
  components of the pre-image of some interval under the function
  $\zeta$ (\ref{zeta}), i.e., in the symmetric case, $F=\sqrt[n]J$
  with $J\subset I$.} but it does not take $A(F)$ to $A(f_t(F))$,
because the mixing takes a neutral product of two Fermi fields in one
component $J_k$ of $F$ to a linear combination of neutral products of
Fermi fields in different components $f_t(J_j)$, belonging to
$C(f_t(F))$ but not to $A(f_t(F))$. Let us call this feature ``charge
splitting'' (stronger than ``mixing'').  

The inclusion situation (\ref{incl}) does not permit to determine the
vacuum modular flow of $A(E)$ from that of $C(E)$, because there is no
vacuum-preserving conditional expectations $C(E)\to A(E)$ that would
imply that the modular group restricts. (Of course, this would be a
contradiction, because we have already seen that the modular group of
$F(E)$, and hence that of $C(E)$, does not preserve $A(E)$.)
Similarly, we cannot conclude that the vacuum modular flow of 
$\widehat A(E)$ should extend that of $C(E)$, or that of $A(E)$. 
Prop.\ 6 below actually shows that this scenario must be {\em excluded}. 

\medskip

\noindent {\bf Application to BCFT}

It is instructive to discuss the consequence of the free Fermi
field mixing and the ensuing charge splitting for $C(E)$ under the
geometric re-interpreta\-tion of boundary CFT, as in Sect.\ \ref{sec:bcft}. 
For definiteness and simplicity, we consider the case when $A$ is the
even subnet of the {\em real}\/ free Fermi net, i.e., $A$ is the
Virasoro net with $c=\frac 12$. Unlike the $c=1$ free Bose net, this
model is completely rational. 

The same considerations as in the previous argument
apply also in this case: Again, the inclusions $A(E)\subset
C(E):=\eps(F(E))\equiv F(E)^{\ZZ_2} \subset \widehat A(E)$ are strict,
the latter because charge transporters for the Ramond sector (weight
$h=\frac1{16}$) do not belong to $C(E)$. The vacuum modular flow for
$C(E)$ is induced by that for $F(E)$, but it does not pass to $A(E)$
or $\widehat A(E)$.

Let therefore $E\subset \Sdot$ be 2-intervals and 
$O=I^\RR_+\times I^\RR_-\subset M_+$ the associated double cones. The
net 
\be
  O\mapsto C(O) =F(E)^{\ZZ_2}
\ee
is a BCFT net intermediate between the ``minimal'' net $A_+(O)=A(E)$
and the ``maximal'' (Haag dual) net $B_+(O)=\widehat A(E)$, see \cite{LR04}. 
It is generated by fields
$\prod_{i=1}^n\psi(u_i)\prod_{j=1}^m\psi(v_j)$ with $n+m=$ even, and
$u_i$ smeared in $I^\RR_+$, $v_j$ smeared in $I^\RR_-$. 

The vacuum modular flow of $C(O)$ mixes $f_tu_i$ with $f_tu_i'$ and
$f_tv_j$ with $f_tv_j'$, where $u\mapsto u'$ and $v\mapsto v'$ are the 
bijections of the two intervals onto each other connecting the two
pre-images of the uniformization function $\zeta$. Hence, if
$\psi(u)^n\psi(v)^m$ (in schematical notation) belongs to $C(Q)$ for a
double cone $Q\subset O$, the vacuum modular flow takes it to linear
combinations of  
\be
  \psi(f_tu)^{n_1}\psi(f_tu')^{n_2}\psi(f_tv)^{m_1}\psi(f_tv')^{m_2}
\ee
with $n_1+n_2=n$, $m_1+m_2=m$. Grouping the charged factors to neutral
(even) ``bi-localized'' products, these generators belong to the local
algebra of 6 double cones $\bigvee_{\alpha=1}^6 C(f_tQ_\alpha)\subset
C(f_t\widehat Q)$ around 6 points as indicated in Fig.\ 3.  

\begin{figure}[htb*]

\hskip45mm \epsfig{file=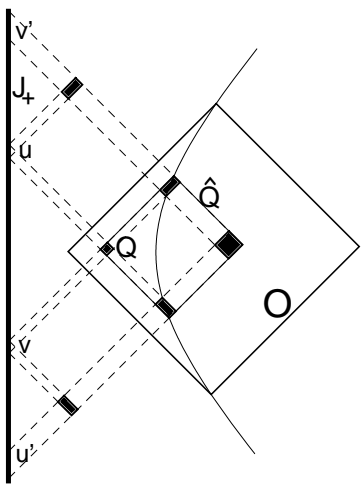,width=50mm} 

\caption {\small The 6 regions mixed by the vacuum modular flow in
  boundary CFT. $(u,v)$ is a point in $Q\subset O$. The boost is the
  distinguished orbit in $O$ as in Sect.\ \ref{sec:modtemp}, and
  defines $u'=-\frac 1u$ and $v'=-\frac 1v$. If $(u,v)$ lies on the
  boost, then the points $(v,u')$ and $(v',u)$ lie on the
  boundary. Consequently, if a double cone $Q\subset O$ around $(u,v)$
  intersects the distinguished orbit, then four of the 6 associated
  double cones $Q_\alpha$ merge with each other, while the other two
  touch the boundary and degenerate to left wedges. (The flow $f_t$
  itself, as in Fig.\ 2, is suppressed.) }    

\end{figure}

In spite of the fact that two of the 6 double cones $Q_\alpha$ lie
outside $\widehat Q$, the corresponding algebras $C(Q_\alpha)$ are
contained in $C(\widehat Q)$. But their bi-localized generators, such
as $\psi(u)\psi(v')$, cannot be associated with {\em points} in
$\widehat Q$, because on the boundary they are localized in the entire
interval $J_+$ spanned by $u$ and $v'$ \cite{LR09}, hence belong 
to $\bigcap_{J_-}C(J_+\times J_-)\subset C(\widehat Q)$. Therefore, in
the geometric re-interpretation of boundary CFT, the discrete mixing
(charge splitting) on top of the geometric modular action induces a  
truely ``fuzzy'' action on BCFT algebras associated with double cones
$Q\subset O$! The fuzzyness seems, however, not to be described by a
pseudo differential operator, as suggested in \cite{SW,TS}, but rather 
reflects the nonlocality of an operator product expansion for
bi-localized fields.

\subsection{Preliminaries for a general theory}

Also in the general case of a local chiral net $A$, there is a notion
of ``charge splitting'': Superselection sectors are described by DHR
endomorphisms of the local net, which are localized in some interval
\cite{DHR}. Intertwiners that change the interval of localization 
(charge transporters) are observables, i.e., they do not carry a
charge themselves, but they may be regarded as operators that
annihilate a charge in one interval and create the same charge in
another interval. These charge transporters do not belong to $A(E)$,
but together with $A(E)$ generate $\widehat A(E)$, see the discussion in
\cite{KLM}. Therefore, one may speculate whether the combination of
geometric action with charge splitting could be a general feature for
the vacuum modular group of suitable $n$-interval algebras
intermediate between $A(E)$ and $\widehat A(E)$, i.e., the modular
group does not preserve the subalgebras $A(F)$, let alone the algebras
of the component intervals $A(J_k)$.   

The discussion of the algebras $A(E)\subset C(E)\subset
\widehat A(E)$ in the preceding subsection shows that there cannot be a
simple general answer. Nevertheless, we can derive a few first general
results.  

\medskip

{\bf Proposition 6: \sl Let $\Phi\in\HH$ be a joint cyclic and
  separating vector for $A(E)$ and $A(E')$, e.g., the vacuum. 

(i) If the modular automorphism group of $(\widehat A(E),\Phi)$ globally
  preserves the subalgebra $A(E)$, then $A(E)=\widehat A(E)$.

(ii) If the adjoint action of the modular unitaries $\Delta^{it}$ for
  $(A(E),\Phi)$ globally preserves $\widehat A(E)$, or, equivalently,
  $A(E')$ then $A(E)=\widehat A(E)$. 

\medskip

\it Proof:} By assumption, $\Phi$ is also cyclic and separating 
for $\widehat A(E)=A(E')'$ and $\widehat A(E')=A(E)'$. Then (i)
follows directly by Takesaki's Theorem \cite[Chap.\ IX, Thm.\ 4.2]{T}. 
For (ii), note that $\Delta^{it}$ preserves $A(E')$ if and only if it
preserves $A(E')'=\widehat A(E)$; and $\Delta^{-it}$ implements the
modular automorphism group for $(A(E)'=\widehat A(E'),\Phi)$. Thus,
the statement is equivalent to (i), with $E$ replaced by $E'$. 
\QED

The obvious relevance of (ii) of Prop.\ 6 is that in the generic case
when $\widehat A(E)$ is strictly larger than $A(E)$, there can be no
vector state satisfying the Reeh-Schlieder property such that $A(E)$
has geometric modular action on $A(E)$ {\em and} on $A(E')$. In
particular, the modular unitaries will not belong to the
diffeomorphism group, but we may expect that Connes spatial
derivatives as in Prop.\ 2 do.  

Recall that we have already seen (in the remark after (\ref{mixdiff}))
that mixing necessarily occurs. By (i) of Prop.\ 6 it is not possible 
that $\widehat A(E)$ has geometric modular action without charge splitting.  

\section{Loose ends}
\label{sec:loose}

We have put into relation and contrasted the two facts that 

(i) in diffeomorphism covariant conformal quantum field theory there
is a construction of states on the von Neumann algebras of local
observables associated with disconnected unions of $n$ intervals
($n$-intervals), such that the modular group acts by diffeomorphisms
of the intervals \cite{KL05}, and  

(ii) in the theory of free chiral Fermi fields, the modular action of
the vacuum state on $n$-interval algebras is given by a combination of
a geometric flow with a ``mixing'' among the intervals \cite{CH}.  

The absence of the mixing in (i) can be ascribed to the choice of
``product'' states in which quantum correlations across different
intervals are suppressed. (In the re-interpretation of 2-interval
algebras as double cone algebras in boundary conformal field theory
\cite{LR04}, the influence of the boundary was shown to weaken -- as
expected on physical grounds -- in the limit when the double cone is
far away from the boundary \cite{LR09}. Indeed, it can be seen from
the formula (\ref{angle}) for the mixing angle that in this limit the
mixing in (ii) also disappears.) On the other hand, there is some
freedom in the choice of product states, which allows to deform the
geometric modular flow within each of the intervals. It comes
therefore as a certain surprise that the geometric part of the vacuum
modular flow in (ii) coincides with the purely geometric flow in the
product states in (i), precisely when the latter are chosen in a
``canonical'' way (involving the simple function $z\mapsto z^n$ on the
circle, corresponding to $\nu=1$ in (\ref{nu-sol}), in the case of
symmetric  $n$-intervals, and the function $g$ (\ref{g}) in the
general case). This means that the relative Connes cocycle  between
the vacuum state and the ``canonical'' product state is just the
mixing, while for all other product states, it will also involve a
geometric component.   

Two circles of questions arise: 

First, is the geometric part of the
vacuum flow specific for the free Fermi model, or is it universal?
And if it is universal, what takes the place of the mixing in the
general case? Putting aside some technical complications of the proof,
the authors of \cite{CH} claim a universal behaviour for free
fields, while in this paper, we have given first indications how the
geometric behaviour should ``propagate'' to subtheories and to field
extensions, also strongly supporting the idea of a universal
behaviour. Insight from the theory of superselection sectors suggests
that the mixing in the general case should be replaced by a ``charge
splitting''. On the other hand, Takesaki's Theorem poses obstructions
against the idea that charge splitting on top of a geometric modular
flow could be the general answer (Prop.\ 6). 

Second, the notion of ``canonical'' ($\nu$=1) in the above should be
given a physical meaning, related to the absence of a geometric
component in the Connes cocycle. In the free Fermi case, the geometric
part of the modular Hamiltonian contains the stress-energy tensor 
$\sim\psi(x)\partial_x\psi(x)$, while the mixing part can be expressed
in terms of $\psi(x_k)\psi(x_l)$ with $x_k$ and $x_l$ belonging to
different intervals. The absence of derivatives suggests that the
Connes cocycle is ``more regular in the UV'' in the case when the
geometric parts coincide, than in the general case. The same should
be true for the generalized product state constructed in Sect.\
\ref{sec:genprod}. A precise formulation of this UV regularity is
wanted. 

\medskip

\noindent{\bf Acknowledgments:} We thank Jakob Yngvason for bringing
to our attention the article of Casini and Huerta \cite{CH}, and
Horacio Casini for discussions about their work. We also thank the 
Erwin-Schr\"odinger-Institute (Vienna) for the hospitality at the 
``Operator Algebras and Conformal Field Theory'' program,
August--December 2008, where this work has been initiated.

\end{document}